**Learning Progression-Guided AI Evaluation of Scientific Models To Support Diverse Multi-Modal Understanding in NGSS Classroom**

*Leonora Kaldaras, Tingting Li, Prudence Djagba, Kevin Haudek, Joseph Krajcik*
*1. CREATE for STEM at Michigan State University; 2. Texas Tech University College of Education*

## Abstract

Learning Progressions (LPs) can help adjust instruction to individual learners' needs if the LPs reflect diverse ways of thinking about a construct being measured, and if the LP-aligned assessments meaningfully measure this diversity. The process of doing science is inherently multi-modal with scientists utilizing drawings, writing and other modalities to explain phenomena. Thus, fostering deep science understanding requires supporting students in using multiple modalities when explaining phenomena. We build on a validated NGSS-aligned multi-modal LP reflecting diverse ways of modeling and explaining electrostatic phenomena and associated assessments. We focus on students' modeling, an essential practice for building a deep science understanding. Supporting culturally and linguistically diverse students in building modeling skills provides them with an alternative mode of communicating their understanding, essential for equitable science assessment. Machine learning (ML) has been used to score open-ended modeling tasks (e.g., drawings), and short text-based constructed scientific explanations, both of which are time- consuming to score. We use ML to evaluate LP-aligned scientific models and the accompanying short text-based explanations reflecting multi-modal understanding of electrical interactions in high school Physical Science. We show how LP guides the design of personalized ML-driven feedback grounded in the diversity of student thinking on both assessment modes.

## Introduction

Learning Progressions (LPs) reflect increasingly sophisticated ways of thinking about a construct [1-3] and therefore can serve as tools for supporting students in attaining deeper understanding of the construct over time [1-3]. LPs can help adjust instruction to individual learners' needs therefore supporting the opportunity to learn [4]. But this is only possible if the LPs reflect diverse ways of thinking about a construct being measured, and if the LP-aligned assessments meaningfully measure this diversity at various levels of sophistication. Further, the process of doing science is inherently multi-modal with scientists utilizing drawings and writings and other modalities to explain phenomena [5]. Thus, fostering deep science understanding requires supporting students in using multiple modalities at various levels of sophistication described by an LP when explaining phenomena. Fostering such understanding therefore requires timely and cognitively appropriate feedback tailored to the diversity of student thinking to be provided to students. To provide such feedback, we first need to evaluate student reasoning with respect to LP levels across different modalities. However, evaluating complex student reasoning in STEM across different modalities and at different sophistication levels often represented by constructed responses (CRs) is time and resource consuming, and requires technologically innovative solutions [6]. Artificial Intelligence (AI) technology, such as machine learning (ML) has shown promise in tackling this challenge. Specifically, ML  has been used to score open-ended modeling tasks (e.g., drawings) [7], and short text-based constructed scientific explanations [8], both of which are time-consuming to score. However, aligning both modalities to provide cognitively appropriate feedback at various levels of sophistication still remains a challenge. This study aims to tackle this challenge of evaluating complex student reasoning in STEM according to the levels of previously validated LP and across both modalities.

We build on a validated NGSS-aligned multi-modal LP reflecting diverse ways of modeling and explaining electrostatic phenomena [9-11] and associated assessments. We focus on students' modeling, an essential practice for building a deep science understanding [1]. Supporting culturally and linguistically diverse students in building modeling skills provides them with an alternative mode of communicating their understanding, essential for equitable science assessment [12]. In the current study we demonstrate using previously validated LP to guide training of ML algorithms to evaluate LP-aligned scientific models (drawings) and the accompanying short text-based explanations  reflecting multi-modal understanding of electrical interactions in high school Physical Science. We address the following research question: *How can a validated  LP guide an ML algorithm training to evaluate LP-aligned multi-modal assessments measuring complex cognitive constructs in STEM to support diverse multimodal understanding?* To answer this RQ we demonstrate how LP guides the design of personalized ML-driven formative feedback grounded in the diversity of student thinking on both assessment modes. LP guidance provides a roadmap for grounding the formative feedback in empirically based, cognitively appropriate ways of modeling and explaining electrostatic interactions grounded in previously validated LP. Such LP-guided ML algorithm training also allows to capture diverse ways of student thinking in the context of modeling electrostatic phenomena and tailor formative feedback in ways that accounts for and builds on this diversity, reflecting a more equitable and constructive approach to ML training for supporting learning [13, 14].

**Theoretical Framework**

We build on the theoretical framework for LP-guided AI training introduced in Kaldaras, Haudek & Krajcik [13]. The framework demonstrates using LPs to guide training for a wide range of AI algorithms focused on preparing AI to evaluate complex student reasoning at various levels of sophistication. The framework's utility has been demonstrated in the context of evaluating LP-aligned constructed responses reflecting stand-along scientific explanations [7], scientific models [8] and math-science sensemaking [15]. The current study demonstrates using the framework to guide training of ML algorithms to evaluate LP-aligned multi- modal assessments combining both scientific models (drawings) and short accompanying explanations.

The study builds on a validated NGSS-aligned LP that integrates the DCIs (qualitative Coulomb's law relationships and charge transfer), CCC of cause and effect, and the SEPs of developing and using models (M) and constructing explanations of (E) electrostatic phenomena. The LP is shown in Table 1.

*Table 1. NGSS-aligned learning progression for electrical interactions [9-11].*

| |
|---|
| **Level 3:** Scientific models and explanations reflect causal relationships that integrate ideas of energy and Coulombic interactions (qualitative, no formula) and charge transfer at the atomic-molecular level to explain electrostatic phenomena. |
| **Level 2:** Scientific models and explanations represent causal relationships that use but don't integrate (or inaccurately integrate) ideas of energy and/or Coulombic interactions (qualitative) and charge transfer at the macroscopic or partially atomic-molecular level to explain electrostatic phenomena. |
| **Level 1:** Scientific models and explanations represent partially causal relationships that use ideas of Coulombic interactions (qualitative), charge transfer and/or energy with inaccurate/incomplete ideas to explain phenomena. |
| **Level 0:** Scientific models and explanations don't represent causal relationships and use ideas of Coulomb's lawn(qualitative), charge transfer and/or energy with significantly inaccurate and/or incomplete ideas to explain phenomena. |

A key step in AI training to evaluate student reasoning lies in designing rubrics that will yield high human-machine agreement and allow for meaningful evaluation of the validity of the AI-based scores to ensure that the AI algorithms capture the same aspects of student responses as a trained human scorer would [14]. While it is possible to design meaningful rubrics for evaluating various tasks using AI without an available LP, this could considerably diminish the usefulness of the resulting scores in terms of providing construct-specific, cognitively appropriate feedback that will help students develop a deeper understanding of a construct beyond the specific assessment items. LPs and LP-aligned assessments, on the other hand, result in data (student responses) that can be meaningfully interpreted in terms of what students can do based on what they demonstrate in their responses, and what support and feedback they need to transition to the next, and subsequent levels of understanding. The information that guides development of such feedback is reflected in the LP levels that describe what students know and should be able to do at various levels of sophistication [3]. LPs therefore represent an overarching roadmap that helps organize and tailor feedback on a wider range of items to help students develop a deeper understanding of a construct across contexts and assessment scenarios. Such feedback, in turn, could support development of transferable knowledge and skills beyond specific learning contexts, which is an ultimate goal of any educational system [1-2].

In the current study we demonstrate how the LP shown in Table 1 helps guide development of rubrics that yield rich and meaningful sources of data to help us accurately place students on an LP level and determine the types of feedback and support they need to help them move up the levels in the context of modeling and explaining electrostatic phenomena. The LP shown in Table 1 guides development of analytic rubrics for both the model (drawing) and the written explanation parts of the assessment item. The same rubric is used for human and ML evaluations of student responses.

The process of LP-guided analytic rubric development is shown in Figure 1. Specifically, for each LP-aligned assessment task the LP guides development of analytic rubric categories that reflect presence or absence of specific ideas that are relevant for capturing student proficiency in the construct described by the LP. Presence of the corresponding ideas is scored as 1, while absence as a 0. One can design as many analytic rubric categories as needed to capture proficiency in a given assessment item. Further, the combinations of "0" and "1"scores for all analytic rubric categories reflects the overall level of student response with respect to the LP. Each analytic rubric category combination can be mapped to a specific LP level. Each combination also reflects specific ideas present or absent in a given response, which allows to tailor feedback to student's specific LP level and the specific information present in student response.Furhter, analytic rubric categories can also be developed to capture specific inaccuracies or incomplete/vague ideas present in student responses. Capturing those ideas can help further personalize and tailor feedback to diverse ways of thinking. The methods section demonstrates how this process was used to develop analytic rubrics and evaluate student models and explanations for the assessment item used in this study.

*Figure 1. LP-guided analytic rubric development for human and automatic scoring.*

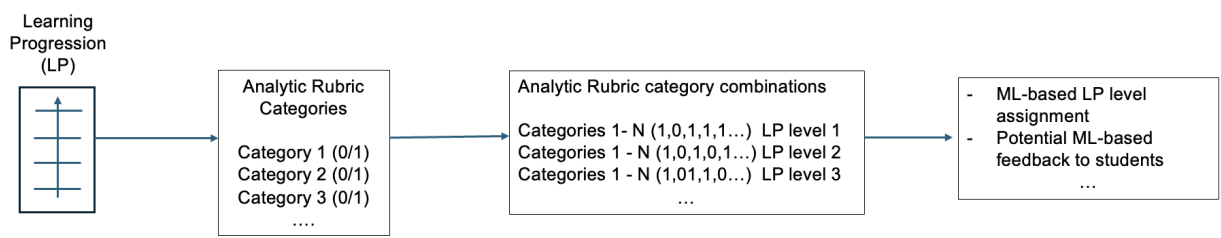

## Methods

*Analytic Rubric Development for Modeling and Explanation Modalities*

We begin with a previously developed item (Figure 1) aligned to LP levels 0-2 (Table 1). The item focuses on ideas of qualitative Coulombic interactions and charge transfer only, without assessing student understanding of energy. The item models interactions between the electroscope parts and a charged rod. The analytic rubric allows to identify presence and absence of essential components of models and explanations therefore permitting developing LP-aligned, feedback tailored to specific student responses for both modalities. The analytic rubric for models and explanations is shown in Table 2.

*Figure 1. Electroscope modeling item.*

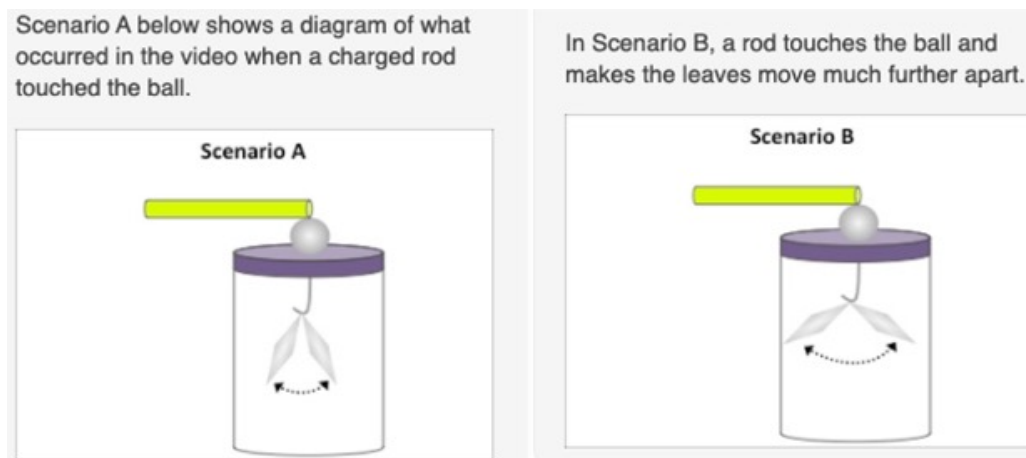

**Question: What is different about scenario A and Scenario B? Justify your answer.**

The modeling rubric contained a total of 13 analytic categories, ten of which reflect accurate components of models that should be present in student responses (categories 1-10 in Table 2). The final LP level assignment reflects student proficiency in developing a causal model explaining the difference between scenario A and B using qualitative Coulomb's law relationships (more charge is associated with larger magnitude of electric force) and charge transfer. To do this, we specified the necessary model components, the relationships between them and the connection to the phenomenon. Briefly, a complete and accurate model should include point charges on all parts of the electroscope (sphere, hook, leaves) as well as the rod. Presence of these model components will indicate student proficiency in using ideas of charge transfer to model how neutral objects (electroscope in this case) become charged through contact with charged objects (rod in this case). In addition there should be more charge on all parts of the model in scenario B compared to A. Finally, student models should show larger repulsive force between leaves of the electroscope in scenario B compared to A. Presents of these components will indicate student proficiency in using qualitative Coulombic relationships, specifically relating the amount of charge on interacting objects to the amount of associated electric force between the objects.

Further, to tackle variability in responses and ensure that resulting feedback can be tailored to the diversity of student thinking, it is important to also capture potential inaccuracies in student models. To do this, we looked at multiple student models and identified 3 broad categories reflecting inaccurate and/or incomplete ideas which we incorporated into the modeling analytic rubric. These categories are described in Table 2 under categories 11-13. Briefly, sometimes student models show the same type of charge on some or all parts of the model (electroscope sphere, hook, leaves). Category 11 aims to capture this inaccuracy. Further, some models show a similar amount of charge on some parts of the electroscope in scenario B compared to A, which could indicate that they don't fully understand that there is more charge transferred in B compared to A. Category 12 aims to capture this inaccuracy. Finally, sometimes students think that the electroscope in part A is not charged at all, and they don't show any charge on the electroscope in part A. Category 13 aims to capture that inaccuracy.

Further, we developed 8  analytic rubric categories for the explanation part of the electroscope item (categories 14-21 in Table 2). Specifically, categories 14 -18 aim to capture accurate ideas that should be present in student responses. Generally, we expect that students will relate the amount of charge transferred from the rod in scenarios A and B to the magnitude of the repulsive force between the electroscope leaves in their explanations. Categories 14 and 15 reflect one component of this causal statement: category 14 captures whether students recognize that scenario A has more charge than scenario B, while category 15 captures whether students recognize that the magnitude of electric force is stronger in B compared to A. Category 16, on the other hand, reflects both of these causal statement capturing that students can relate the amount of charge to the resulting magnitude of electric force when comparing scenarios A and B. Further, categories 17 and 18 capture presence of other important accurate ideas, like charge transfer (category 17) and fundamental property of charges (category 18). Like the modeling rubric, we wanted to identify common inaccurate or incomplete ideas in student explanations (shown as categories 19-21). Specifically, category 19 captures whether students think that the electroscope in part A is not charged at all, which is similar to category 13 for the modeling modality. Categories 20 and 21 aim to capture incomplete explanations reflected in students describing their observations with no use of disciplinary ideas (category 20) or lack of comparison between scenarios (category 21).

*Table 2. Analytic rubric for the model and explanation component of the electroscope item.*

| Category | Description |
|---|---|
| 1 | Point charge (either + or – ) on the rod in scenario A |
| 2 | Point charge on the metal ball. The charge must be the same type as shown in the rod in scenario A. *Alternatively, models can show charge transfer from the rod to the ball with arrows, and not explicitly show point charges on the ball (there should be charges on the rod)* |
| 3 | Point charge on the hook of the electroscope. The charge must be the same type as shown on the rod in scenario A. *Alternatively, models can show charge transfer from the ball to the hook/foil leaves with arrows, and not explicitly show point charges on the hook (there should be charges on the ball)* |
| 4 | Point Charge on the leaves of the electroscope in scenario A. The charge must be the same type as shown in the rod in scenario A. |
| 5 | Clearly indicates repulsive Electric force causes leaves to move, by using arrows or force representations and pointing in opposite directions between the leaves in scenario A |
| 6 | Point charge on the rod in scenario B. The charge must be the same type as shown on the rod in scenario A. There must be more point charges on the rod in scenario B than in scenario A. |
| 7 | Point charge on the sphere of the dome in scenario B. The charge must be the same type as shown on the sphere of the dome in scenario A. There must be more point charges on the sphere in scenario B than in scenario A. *Alternatively, models can show charge transfer from the rod to the ball with arrows, and not explicitly show point charges on the ball* |
| 8 | Point charge on the hook of the electroscope in scenario B. The charge must be the same type as shown on the hook in scenario A. There must be more point charges on the hook in scenario B than in scenario A. *Alternatively, models can show charge transfer from the ball to the hook with arrows, and not explicitly show point charges on the hook* |
| 9 | Point Charge on the leaves of the electroscope in scenario B. The charge must be the same type as shown in the leaves in scenario A. A. There must be more point charges on the leaves in scenario B than in scenario A. |
| 10 | Clearly indicates repulsive Electric force causes leaves to move, by using arrows or force representations and pointing in opposite directions between the leaves in scenario B.. The repulsive arrows should be bigger or bolder (or both) for scenario B than for scenario A. |
| 11 | Model shows both types of charges on one or more part of the electroscope in one or both scenarios. *This can be ignored if positive and negative charges are not accumulated in specific locations.* |
| 12 | Similar amount of charge on one or more parts of the electroscope in scenario A and B. This category only applies if they show the same type of charge through the entire model. |

| 13 | Either the rod in scenario A is not charged or the whole electroscope are not charged in scenario A |
|---|---|
| 14 | States that:<br>- rod in scenario B has more charge OR<br>- rod in scenario A has less charge OR<br>- student can state that scenario A has less charge than Scenario B |
| 15 | States that repulsive electric force or electric field is:<br>-stronger in scenario B than in scenario A<br>-weaker in scenario A than in scenario B |
| 16 | Relate the amount of charge to the magnitude of the repulsive electric force in both scenarios. States that:<br>-Larger amount of charge in scenario B compared to scenario A causes stronger repulsive force (**or** causes the leaves to move apart more)<br>-Smaller amount of charge in scenario A compared to scenario B causes weaker repulsive force (**or** causes leaves to move apart less) |
| 17 | States that Rod/parts of the system transfers charge to the foil leaves or any part of the electroscope in one or both scenarios. No comparison between scenario A and B is necessary for this category |
| 18 | States that similar charges repel |
| 19 | -States that the rod (or any other part of the electroscope) is neutral (not charged) in Scenario A but charged in scenario B;<br>**OR**<br>-states that the electroscope leaves are neutral in A but charges in B<br>**OR**<br>-States that the charged rod is not transferring any charge to the electroscope (metal ball, hook, foil leaves) or foil leaves in scenario A |
| 20 | Description of event only (no causality implied or disciplinary idea used) |
| 21 | Does not develop a **comparison** response **including both scenarios** explaining why foil leaves mover further away in B compared to A |

Table 3 describes alignment analytic rubric combinations for modeling and explanation modality at each LP level. In this study we evaluate models and explanations separately and provide an LP level assignment for a model and an explanation. Therefore, the corresponding potential feedback is provided separately for a model and an explanation.

*Table 3. Analytic rubric Categories combinations alignment with LP levels.*

<u>Level 2</u>: *models* show charges on almost all electroscope parts (rod, sphere, hook, leaves, missing on no more than 2 parts is permissible), more charge transferred from the rod to all electroscope parts in scenario B, and a greater repulsive force in scenario B (score of "1" in at least 8 of total ten categories (categories 1-10), score of "0" in categories 11-13). *Explanations* provide a causal statement relating the difference in the amount of charge on the rod in both scenarios to the amount of charge transferred to the foil leaves and the resulting magnitude of the repulsive force (score of "1" for category 16). Additional accurate information is permissible (score of "1" in categories 17 and 18), no inaccuracies (score of "0" in categories 19-21).

Level 1: *models* miss more than 2 components (e.g., charge on more than one electroscope part, repulsive force indicator etc.) but no more than 4 components (score of "1" in at least 6 of total ten categories (categories 1-10)). *Explanation* only contains one component of the causal statement. Inaccuracies in both modalities are permissible (score of "1" on category 14 or 15). Inaccuracy categories are permissible for both model (score of "1" on categories 11-13) and explanation (score of "1" on categories 19-21).

Level 0: models show charges on less than 6 components of the model (score of 1 on less than 6 of the total ten categories (categories 1-10). *Explanation* is absent or only contains inaccuracies (score of "1" on categories 11-13).

*Data Sources*

The Electroscope item was administered to 9th-grade students participating in the NGSS-aligned curriculum study. Unit 1 focused on ideas related to Coulomb's law as related to electrical interactions. The Electroscope item was administered as part of the Unit 1 pre and post-test and student responses from the posttest were used for the analysis reported here.

*Human Scoring*

We coded ~200 randomly selected student models and accompanying explanations to ensure that the rubrics for both modalities were easy to use and applied to a range of responses. The rubrics and the coded responses were then reviewed by the researchers in the group. Clarifications of rubric criteria and necessary additions were made to ensure the usability of the rubric. Three undergraduates were trained to apply the rubric to student responses. Training was done in subsets of several hundred responses and coded independently by coders.

*ML Model Training and Testing*
<u>*Modeling Modality*</u>

We used supervised ML, specifically convolutional neural network analysis approach with Res-Net18 architecture as feature extraction network [16]. The training data set contained 884 responses (73 % of the overall data set) and the testing set contained 327 images (27% of the dataset). During training, we use the pretrained ResNet-18 (Residual Network) architecture, modifying its final fully connected layer to deliver binary output for our classification needs. The ResNet-18 architecture, noted for its deep residual learning framework, was employed as our feature extraction network [16]. This network, with its depth of layers and residual connections, is particularly adept at learning from small datasets, which often pose challenges for deep learning models due to the risk of overfitting [16]. To accommodate the input dimensionality and maintain consistency with the ResNet architecture, we set $d = 512$ (feature dimensionality) and resized all images to $W = H = 224$ (pixels).

Our model was implemented in PyTorch, benefitting from its flexible programming environment and efficient computational graph dynamics [17]. Optimization during training was conducted using the Adam optimizer, with a learning rate of $1e$ - 4, balancing the advantages of adaptive gradient methods with the need for precision in the weight update process [18]. An NVIDIA GeForce GTX 1080Ti graphics card expedited the training process, enabling the efficient optimization of the model. Throughout the cross-validation process, we systematically assessed and saved the best-performing models according to validation metrics, opting for F1 score or accuracy based on the dataset's balance.

We tested multiple data augmentation approaches to ensure best human-machine agreement for training and testing stages. A detailed discussion of performance for various augmentation approaches has been recently published [19]. In this paper we report results of human-machine agreement achieved using the SMOTE augmentation approach which yielded the highest agreement.

*Explanation Modality*

We used a supervised deep learning approach utilizing the Bidirectional Encoder Representations from Transformers (BERT) model to classify responses across categories 14 to 21. Specifically, we used the bert-base-uncased model as the foundation for our feature extraction network. The dataset consisted of textual justifications mapped to multiple categories, with missing values in both the justification and categorical fields handled by appropriate imputation strategies. The training dataset contained an 80-20 split for training and testing purposes on the 1060 students response.

During preprocessing, textual inputs were tokenized using the BERT tokenizer with a maximum sequence length of 128 tokens to ensure consistency across input representations. The training pipeline involved encoding textual data, which was subsequently passed through the BERT model. The textual representations were extracted through the pooler_output layer of BERT, which serves as an encoder for the texts. The final classification layers consisted of additional dense layers with ReLU activation, followed by a sigmoid-activated output layer to accommodate multi-label classification across the eight categories.

The model was implemented using TensorFlow and trained with the Adam optimizer, adopting a learning rate of 2e-5 to balance convergence speed and generalization. To mitigate overfitting, we introduced dropout layers with a 30% dropout rate and utilized early stopping based on validation loss, ensuring that the optimal model was retained. The model was trained for up to 10 epochs with a batch size of 16.

**Results**

*Human Scoring*

The codes coded models and explanations separately. Results from independent coding on subsets were checked for IRR (Krippendorff, 2004). We used a threshold of Krippendorff's alpha greater than 0.8 between human coders for each analytic category [20]. We then checked for human IRR. Categories that showed <0.8 Krippendorf's alpha between coders were discussed by the coders until agreed upon and the rubric was updated. A total of 1211 modeling and explanation responses from students collected in 9th grade Physical Science classroom were scored by trained human scorers. This data set is subsequently used to train the ML model.

*ML Scoring*

Precision, recall, and F1 score are key evaluation metrics used in classification tasks to measure the performance of a model. Precision refers to the proportion of correctly predicted positive instances out of all instances predicted as positive. It indicates how often the model's positive predictions are actually correct. Recall (also known as sensitivity) measures the proportion of actual positive instances that the model correctly identifies, showing how well the model captures all relevant instances. F1 score is the harmonic mean of precision and recall, providing a balanced measure that accounts for both false positives and false negatives. See [21,22]. We used these measures to evaluate ML model performance for modeling and explanation modalities as discussed further.

<u>*ML Scoring for Models*</u>

Table 5 shows the final human-machine agreement for each scoring category for the training and testing stages. As shown in Table 5, human-machine agreement for all categories for the testing stage is above 90% accuracy, reflecting very high agreement. Other measures such as precision, recall and F-1 score are also above 0.9 indicating very good model performance. Further, accuracy for the testing stage is also above 90% accuracy, reflecting very high agreement. Other measures such as precision, recall and F-1 score are also above 0.8 for most categories indicating good model performance. We note that some of the categories with the lowest performance metrics such as F-1 score for the training stage are rubric categories associated with inaccuracies - categories 11-13. Notice also that these categories have a small overall number of positive cases available in the dataset as shown in Table 6. Similarly, a somewhat lower performing category (but still within acceptable range)-category 7 also has a lower number of positive cases in the dataset as shown in Table 6. Overall, this data suggests that the supervised ML approach accurately detected the model components and critical relationships within the model that were outlined in the rubric for training and testing stages.

*Table 5. Human-Machine agreement using CNN algorithm performance with SMOTE augmentation for the modeling modality.*

| Category | Training Stage (cross validation) | | | | | Testing Stage | | | | |
|---|---|---|---|---|---|---|---|---|---|---|
| | accuracy | 95% CI | precision | recall | F1 score | accuracy | 95%CI | precision | recall | F1 score |
| C1 | 0.94 | (0.93, 0.94) | 0.94 | 0.94 | 0.94 | 0.94 | (0.89, 0.99) | 0.94 | 0.92 | 0.93 |
| C2 | 0.96 | (0.95, 0.97) | 0.96 | 0.96 | 0.96 | 0.97 | (0.93, 1.01) | 0.95 | 0.93 | 0.94 |
| C3 | 0.97 | (0.96, 0.97) | 0.97 | 0.97 | 0.97 | 0.97 | (0.93, 1.00) | 0.90 | 0.94 | 0.93 |
| C4 | 0.95 | (0.94, 0.95) | 0.95 | 0.95 | 0.95 | 0.93 | (0.89, 0.98) | 0.90 | 0.90 | 0.90 |
| C5 | 0.96 | (0.95, 0.96) | 0.95 | 0.95 | 0.95 | 0.96 | (0.91, 1.00) | 0.94 | 0.94 | 0.94 |
| C6 | 0.91 | (0.90, 0.92) | 0.91 | 0.91 | 0.91 | 0.91 | (0.87, 0.96) | 0.90 | 0.84 | 0.87 |
| C7 | 0.95 | (0.94, 0.95) | 0.95 | 0.95 | 0.95 | 0.94 | (0.91, 0.97) | 0.97 | 0.64 | 0.71 |
| C8 | 0.95 | (0.94, 0.96) | 0.96 | 0.95 | 0.95 | 0.94 | (0.90, 0.96) | 0.79 | 0.86 | 0.82 |
| C9 | 0.94 | (0.93, 0.95) | 0.94 | 0.94 | 0.94 | 0.93 | (0.89, 0.97) | 0.90 | 0.80 | 0.84 |
| C10 | 0.95 | (0,94, 0.96) | 0.95 | 0.95 | 0.95 | 0.95 | (0.91, 1.00) | 0.95 | 0.90 | 0.92 |
| C11 | 0.93 | (0.92, 0.93) | 0.93 | 0.93 | 0.92 | 0.91 | (0.88, 0.94) | 0.65 | 0.55 | 0.56 |
| C12 | 0.92 | (0.91, 0.93) | 0.93 | 0.92 | 0.92 | 0.91 | (0.87, 0.94) | 0.73 | 0.65 | 0.68 |
| C13 | 0.96 | (0.95, 0.96) | 0.96 | 0.96 | 0.96 | 0.92 | (0.89, 0.96) | 0.83 | 0.72 | 0.76 |

*Table 6. Percent of positive cases for each scoring category for modeling modality.*

| Category | Percent of positive cases (%) |
|---|---|
| 1 | 33.99 |
| 2 | 16.43 |
| 3 | 11.89 |
| 4 | 19.57 |
| 5 | 21.3 |
| 6 | 23.29 |
| 7 | 11.81 |
| 8 | 8.92 |
| 9 | 17.84 |
| 10 | 20.48 |
| 11 | 8.92 |
| 12 | 8.51 |
| 13 | 8.9 |

## *ML Scoring for Explanations*

Performance evaluation across categories revealed strong human-machine agreement, with validation accuracy exceeding 90% for most categories, as shown in Table 7. The highest performance was observed in categories 15 and 18, where precision reached 91.30% and 100%, respectively. However, certain categories, such as 17 and 19, exhibited lower F1 scores due to class imbalance as shown in Table 8 and inherent challenges in label consistency. Despite these variations, the overall performance suggests that the BERT-based approach effectively captured key relationships in the data and aligned well with human scoring patterns. These findings highlight the effectiveness of BERT in multi-label classification tasks while also emphasizing the need for further refinements in certain categories to enhance recall and balance precision-recall trade-offs. However, these results indicate that this rubric is appropriate for evaluating student explanations on all major accuracy categories (categories 14-16) and potentially inaccuracy categories as well.

*Table 7: Human-machine agreement for the ML training stage of models scoring*

| Category | Testing Stage | | | |
| --- | --- | --- | --- | --- |
| | Accuracy (%) | Precision (%) | Recall (%) | F1 score (%) |
| C14 | 91.70 | 85.29 | 89.23 | 87.21 |
| C15 | 97.56 | 91.30 | 87.50 | 89.36 |
| C16 | 92.19 | 77.27 | 85.00 | 80.95 |
| C17 | 93.65 | 57.14 | 28.57 | 38.09 |
| C18 | 96.58 | 100 | 36.36 | 53.33 |
| C19 | 93.65 | 50.00 | 38.46 | 43.47 |
| C20 | 94.63 | 87.75 | 89.58 | 88.65 |
| C21 | 90.73 | 75.00 | 58.06 | 65.45 |

Table 8. *Percent of positive cases for each scoring category for explanation modality.*

| Assessment of imbalance | Category 14 | Category 15 | Category 16 | Category 17 | Category 18 | Category 19 | Category 20 | Category 21 |
| --- | --- | --- | --- | --- | --- | --- | --- | --- |
| Percent positive cases n=1066 | 33.3 | 10.7 | 19.2 | 8.4 | 5.0 | 7.2 | 23.8 | 12.4 |

*Examples of Output and Potential Feedback*

We further demonstrate some examples of scored models and accompanying explanations and discuss potential feedback that can be tailored to the specific responses based on the scoring and corresponding LP level assignment.

For example, figure 2 shows the model that is consistent with the highest possible LP level for this item- level 2, while the explanation provides a level 1 response. Specifically, notice that the model shows all the necessary components, including all the charges and repulsive forces on all parts on the electroscope in both scenarios, which is consistent with LP level 2 for the modeling modality. On the other hand, the explanation modality only reflects students recognizing that there is more charge in scenario B compared to A without relating it to magnitude of associated electric force, which is needed to attain LP level 2 on this modality. Therefore, the proposed feedback statement recognizes the accuracy of the model (red text), while providing feedback for explanation (blue text) to help the student attain level 2 on this modality by relating the amount of charge to the magnitude of electric force to explain the phenomenon in question.

Further, figure 3 shows sample response consistent with LP level 1 on both modalities with no inaccuracies. Notice that the model misses charge components on the sphere and hook in both scenarios, which reflects level 1 on modeling modality. Further, the accompanying explanation does not relate the difference in the amount of charge to the difference in magnitude of the associated repulsive force- similar to the previous example. Possible feedback addresses both of these shortcomings to help the student attain a higher level for both modalities.

*Figure 2. Sample LP level 2 response and potential feedback.*

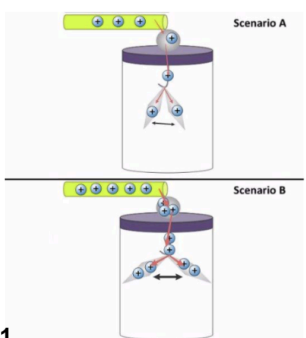

**Category 1 (rod) =1**
**Category 2 (sphere) = 1**
**Category 3 (hook) = 1**
**Category 4 (leaves) = 1**
**Category 5 (repulsive force) = 1**

**Category 6 (rod) =1**
**Category 7 (sphere) = 1**
**Category 8 (hook) = 1**
**Category 9 (leaves) = 1**
**Category 10 (repulsive force) = 1**

Category 11-13: 0

<u>Explanation</u>: In scenario B, the rod has more charge and can transfer more as well while in scenario A, the amount is reduced.
**Category 14-1**
Category 15- 0
Category 16-0
Category 17-0
Category 18-0
Category 19-0
Category 20-0
Category 21-0

Model: Level 2  (maximum possible)          Explanation: Level 1

<u>Possible feedback</u>: your model accurately describes how the difference in the amount of charge on the rod in scenario B compared to A affects the observations. Make sure your explanation describes why bigger charge on the rod in scenario B causes the leaves in scenario B to move further apart.

*Figure 3. Sample LP level 3 response and potential feedback.*

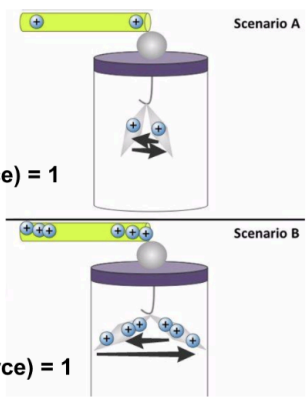

**Category 1 (rod) =1**
Category 2 (sphere) = 0
Category 3 (hook) = 0
**Category 4 (leaves) = 1**
**Category 5 (repulsive force) = 1**

**Category 6 (rod) =1**
Category 7 (sphere) = 0
Category 8 (hook) = 0
**Category 9 (leaves) = 1**
**Category 10 (repulsive force) = 1**

Category 11-13: 0

<u>Explanation</u>: The charged rod in Scenario B has a bigger charge than the rod in Scenario A.
**Category 14-1**
Category 15- 0
Category 16-0
Category 17-0
Category 18-0
Category 19-0
Category 20-0
Category 21-0

Model: Level 1                    Explanation: Level 1

Finally, Figure 4 shows an example of a model reflecting level 0 of the LP on both modalities. Specifically, the model shows both types of charges on the electroscope, which is consistent with inaccuracy category 11, and no accompanying explanation. Notice that the feedback for the modeling modality focuses on recognizing that students showed charges on their model, and pointing out that both types of charges were shown. The feedback also pushes students to think about how charges cause differences in observations and show their understanding on both modeling and explanation modalities.

      These few examples demonstrate how the rubric and the LP-guided approach discussed in this study can be used to tailor feedback to a wide range of student responses reflecting diverse ways of thinking and sophistication. We used this approach to design feedback statements for a wide range of models and accompanying explanations and plan to pilot them.

*Figure 4. Sample LP level 0 response containing inaccuracy in modeling and potential feedback.*

Categories 1-10: 0
**Category 11: 1**
Category 12: 0
Category 13: 0

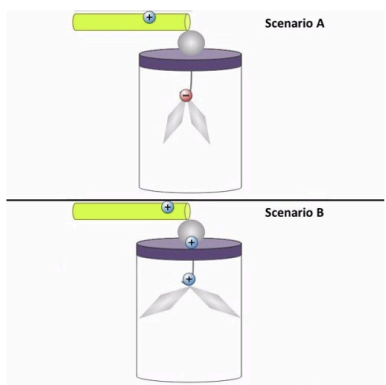

Explanation: none

Categories 14-21: 0

Model: LP level 0                       Explanation: LP level 0

<u>Possible Feedback</u>: Your model shows opposite charges on different parts of the electroscope. Make sure your model explains how the charges to cause the difference in observations of the leaves' motion in scenarios A and B. Provide a brief written explanation of your proposed model. Make sure your explanation includes how the charges affect electroscope leaves in both scenarios to cause the difference in observations.

## Discussion

It is challenging to train AI algorithms to recognize complex reasoning such as that reflected in students' scientific models and explanations. This is because ML algorithms should be trained to go beyond simple features of a given image to recognize specific aspects that are important for the practice of modeling focusing on evaluating causal aspects of scientific models explaining phenomena. Similarly, ML models should be trained to recognize relevant components of scientific explanations. This is especially challenging when we aim to provide cognitively appropriate feedback tailored to the diversity of student thinking reflected in these modalities. The reason is that often scientific models at various LP levels might look very similar (compare level 2 and level 1 models in figures 2 and 3 respectively), but in reality represent qualitatively different levels of understanding. Furthermore, diversity of student ways of modeling and explaining at various levels of sophistication is often integrated with various inaccurate and /or incomplete ways of thinking  (e.g., inaccurate model shown in Figure 4). If ML algorithms are  not able to accurately capture these important differences, then we will not be able to design accurate and targeted LP-aligned feedback, which in turn defies the purpose of using ML techniques  to solve one of the central current problems in education- personalizing education to individual learners' needs. It is therefore important to design approaches that leverage everything we know about how proficiency in a given construct develops, which is reflected in the LP-based vision, when designing AI-based methods for evaluating student learning. The current study demonstrates how an LP can be used to guide ML training to evaluate student thinking on two modalities: models and explanations, both of which are crucial for supporting deep science understanding.

The proposed LP-guided ML training process yields results that are meaningful with respect to LP levels, provide high human-machine agreement on most cases, and allow meaningfully capture the diversity of student thinking on both modalities and tailor formative feedback to individual student needs.

In cases when human-machine agreement is not sufficient, future work will focus on providing more examples for ML training to ensure that ML algorithms have sufficient number of pre-scored responses to learn to recognize specific features in student models and explanations. Notice that insufficient agreement was mostly demonstrated for categories that capture inaccuracies (categories 11-13 for modeling and categories 19 and 21 for explanations). These categories often don't have a sufficient number of positive cases, or represent highly diverse ways of ways which could be characterized in those categories. For example, in the case of modeling modality, categories with the lowest F1 score- categories 11 and 12, both of which have few responses in the dataset as shown in Table 6. Further, category 12 is very diverse because a similar amount of charge can be shown on a wide range of electroscope parts, in both scenarios, all of which would classify the model in this category. Similarly,  category 11 reflects models that show both types of charges, which can also be shown on different parts of the electroscope and in both scenarios, making a range of possible responses highly diverse. This is in contrast to scoring a "1" in categories 1-10, where there is basically only one possible way of attaining that score. Therefore, it is possible that lower human-machine agreement on these categories could be due to the fact that these categories offer a wider range of possible answers that could be classified in that category and smaller number of available responses that don't necessarily reflect this diversity. However, additional empirical studies are needed to further confirm this suggestion.

***Study's Significance***

This process of LP-guided AI algorithm training described here (Figure 1) represents a transparent and principle-based approach for designing LP-aligned, personalized feedback for any constructed response assessments (including scientific models, text-based explanations etc.). Defining analytic categories in this manner allows for easy identification of human-machine misscores by providing a straightforward way to pinpoint specific analytic rubric categories that were misscored. This property has the potential to improve overall validity of the associated AI-based scoring system. Importantly, this LP-driven approach to AI training allows us to go beyond using AI to perform specific tasks (e.g., scoring isolated assessment items) and train AI to guide the learning process in ways that are grounded in relevant cognition theories and foster multimodal understanding beyond specific tasks and modalities.